\documentclass[a4paper,UKenglish,cleveref, autoref, thm-restate]{lipics-v2021}

\usepackage{cite}
\usepackage{amsmath,amssymb,amsfonts}
\usepackage{algorithmic}
\usepackage{graphicx}
\usepackage{textcomp}
\usepackage{hyperref}
\usepackage{xcolor}
\usepackage{tikz}
\usepackage{bbold}
\usepackage[normalem]{ulem}

\nolinenumbers
\def\BibTeX{{\rm B\kern-.05em{\sc i\kern-.025em b}\kern-.08em
    T\kern-.1667em\lower.7ex\hbox{E}\kern-.125emX}}

\newtheorem{property}[theorem]{Property}

\makeatletter
\@fleqnfalse
\@mathmargin\@centering
\makeatother

\newboolean{showcomments}
\setboolean{showcomments}{true}
\ifthenelse{\boolean{showcomments}}
{ \newcommand{\mynote}[3]{
     \fbox{\bfseries\sffamily\scriptsize#1}
       {\small$\blacktriangleright$\textcolor{#3}{{\bf #2}}}$\blacktriangleleft$}}
       { \newcommand{\mynote}[2]{}}

\newcommand{\colorComment}[1]{\textcolor{black}{#1}} 

\title{Byzantine Attacks  Exploiting Penalties in Ethereum PoS
}

\author{Ulysse Pavloff}{Université Paris-Saclay, CEA, LIST \and Palaiseau, France}{ulysse.pavloff@cea.fr}{https://orcid.org/0000-0003-4125-3306}{}
\author{Yackolley Amoussou-Guenou}{Université Paris-Panthéon-Assas, CRED \and Paris, France}{}{https://orcid.org/0000-0002-6942-0412}{}
\author{Sara Tucci-Piergiovanni}{Université Paris-Saclay, CEA, LIST \and Palaiseau, France}{}{https://orcid.org/0000-0001-9738-9021}{}

\authorrunning{U. Pavloff, Y. Amoussou-Genou, and S. Tucci-Piergiovanni} 

\Copyright{Ulysse Pavloff, Yackolley Amoussou-Genou, and Sara Tucci-Piergiovanni} 



\category{} 

\relatedversion{} 

\ArticleNo{24}
\begin{document}

\maketitle

\begin{abstract}
In May 2023, the Ethereum blockchain experienced its first inactivity leak, a mechanism designed to reinstate chain finalization amid persistent network disruptions. This mechanism aims to reduce the voting power of validators who are unreachable within the network, reallocating this power to active validators.
This paper investigates the implications of the inactivity leak on safety within the Ethereum blockchain. 
Our theoretical analysis reveals scenarios where actions by Byzantine validators expedite the finalization of two conflicting branches, and instances where Byzantine validators reach a voting power exceeding the critical safety threshold of one-third. Additionally, we revisit the probabilistic bouncing attack, illustrating how the inactivity leak can result in a probabilistic breach of safety, potentially allowing Byzantine validators to exceed the one-third safety threshold.
Our findings uncover how penalizing inactive nodes can compromise blockchain properties, particularly in the presence of Byzantine validators capable of coordinating actions.
\end{abstract}

\section{Introduction}

Ethereum has transitioned to its proof-of-stake (PoS) protocol in September 2022, making a shift from proof-of-work to the  more energy efficient proof-of-stake. The design of the Ethereum PoS, however, is quite intricate and recent research focused on its formalization, properties, and thorough analyses \cite{neu_ebb_2021, pavloff_ethereum_2023}.
The distinct feature of the Ethereum PoS protocol is the hybridization of classical Byzantine Fault-Tolerant (BFT) Consensus \cite{pbft} within the framework of Nakamoto-style blockchains \cite{ouroboros, bitcoin, ethereum_pow}.

In Nakamoto-style blockchains, each peer maintains a local tree-like data structure. A deterministic rule, often termed the fork-choice rule, selects a chain from this structure, allowing for forks but ultimately reconciling to a single common chain.

In contrast, BFT Consensus blockchains \cite{ Tenderbake, buchman_et_al_2018}  operate without admitting forks in the blockchain. Through explicit voting for block proposals, where participants' votes align with their stake in the network, these blockchains ensure that once a block is added to the chain, it remains permanently incorporated, never revoked due to a fork. In blockchain terms, such a block is \emph{finalized}, signifying settlement of all transactions within it. To establish a finalized chain, BFT Consensus blockchains use Byzantine-tolerant super-majority quorums, assuming a tolerance threshold below one-third of Byzantine voting power. 

In terms of properties, BFT Consensus blokchains are always safe because they do not fork, but can stop growing during network partitions or erratic network behavior. On the other hand, Nakamoto-style blockchains can fork during partitions or erratic network behavior but are always live, because the tree will not stop to grow under these circumstances. 
The trade-off between Safety and Liveness during network partitions is a striking consequence of the CAP theorem \cite{brewer_2010, GL02}. 

Instead, Ethereum PoS uniquely features a finalized chain as the root of a chain that can fork. Chain finalization relies on Byzantine voting quorums, while a fork-chain rule aids validators in resolving forks.
Maintaining both finalized and non-finalized chains within a single data structure, Ethereum PoS strives for a delicate balance between safety and chain growth. Safety within Ethereum asserts the non-forkable nature of the finalized chain, while Liveness guarantees its continued growth, akin to BFT Consensus blockchains. However, in contrast to BFT Consensus blockchains, Ethereum PoS ensures that block addition in the forkable part of the chain remains unimpeded, even during network disruptions or failures \footnote{This attribute, essentially non-blocking Liveness, is denoted as Availability in \cite{pavloff_ethereum_2023}.}.

Nonetheless, finalization procedures, based on Byzantine quorums, still carry the inherent limitation that if more than one-third of the voting power resides with honest validators who are unreachable within the network, the protocol may fail to achieve finalization. These validators are perceived as inactive and do not contribute to finalization.  To mitigate this scenario, thereby restarting finalization in case of long periods of bad network behavior, the \textit{inactivity leak}, was introduced \cite{buterin_casper_2017}. Intuitively, the inactivity leak is a penalty mechanism that erodes the stake of inactive validators (and with that their voting power) to redistribute voting power to active validators. The inactivity leak can be seen as mechanism to restore Liveness during network partitions, which then poses a theoretical risk on Safety by the CAP theorem. Indeed, the initial introduction of the inactivity leak briefly mentions the prospect of conflicting finalized blocks, meaning a loss of Safety, however, it falls short in elucidating the exact events that could lead to these issues and their severity.
This is particularly crucial in the context of potential interference by Byzantine validators, a variable that necessitates closer examination.

We address this question and give the first formal description of the inactivity leak and its impact on the protocol by considering different configurations in terms of initial Byzantine voting power and outcome in terms of Safety loss. We are interested for Safety loss in two distinctive outcomes: (1) the finalization of two conflicting chains, and (2) the break of the Safety threshold, meaning the Byzantine stake proportion of more than one-third. 

Our theoretical analysis aims to pinpoint the significant role the inactivity leak mechanism plays in the Ethereum protocol.
We first assess conflicting finalization when all validators are honest. By leveraging the CAP theorem, we emulate a network partition to establish the time it takes for the network to restore Liveness while breaking Safety because of network regions finalizing independently on different chains. 

We then analyze, considering different Byzantine initial voting power proportions, the time it takes to break Safety under the same network condition and where Byzantine validators can coordinate and are not affected by network partition. We present attacks in which Byzantine validators can either accelerate the time that it takes to break Safety by finalizing conflicting chains or break the Safety threshold acquiring more than one-third of voting power. 
We also revisit a known attack called the Probabilistic Bouncing attack in light of the inactivity leak. The attack was initially identified as a means to delay finalization even during periods of good network conditions. We demonstrate that this attack can be exploited to break the Safety threshold in a probabilistic manner during network synchronous periods, i.e., without partition. 

\colorComment{
Let us note that the conditions required for these attacks to happen and persist, to the point of actually threatening Safety, make them unlikely in real-world settings, primarily due to the time scales involved. Nonetheless, Byzantine validators can arguably cause other validators to lose stake before compromising Safety. Indeed} our work focuses on mechanisms penalizing inactive validators, akin to those found in other PoS blockchains such as Polkadot \cite{wood_polkadot_2016} or Tezos \cite{goodman_tezos_2014}. 
\colorComment{In this respect}, we consider our analysis a crucial endeavor in exploring penalties mechanisms within Byzantine Fault Tolerance (BFT) analysis, aiming to provide valuable insights for future research.

The paper is organised as follows:  we first present a description of the system model and blockchain properties (\autoref{sec:sysMod-blockProp}), prerequisites for understanding the following section detailing Ethereum's protocol (\autoref{sec:ethProtocol}). Subsequently, we delve into the formalization of the inactivity leak (\autoref{sec:inactivityLeak}) and go through the aforementioned scenarios(\autoref{sec:analysis}) for our analysis. Then comes the related work (\autoref{sec:relatedWork}) before reaching our conclusions.

\section{System Model \& Blockchain Properties}\label{sec:sysMod-blockProp}

We consider a system composed of a finite set $\Pi$ of processes called \emph{validators}.
There are a total of $n$ validators. Each validator owns a \emph{stake}. The stake refers to the amount of cryptocurrency (ETH) owned by each validator, serving as a metric of their influence in the consensus protocol (cf. \autoref{sec:ethProtocol}).
Validators possess a unique public/private key pair used for cryptographic signing, and can be identified by their public key.
We assume that digital signatures cannot be forged. 
Validators have synchronized clocks\footnote{Clocks can be offset by at most $\tau$,
this way, the offset can be captured as part of the network delay.}. 
Time is measured by periods of 12 seconds called \emph{slots}, a period of 32 slots is called an \emph{epoch} which serves as the largest time unit in the protocol. Throughout the remainder of this paper, we employ the term ``proportion'' concerning a validator set to denote the ratio of their combined stake to the total value staked. Initially capped at 32 ETH, the stake of individual validators has the potential to decrease.

\paragraph*{Network} Validators communicate by message passing.
We assume the existence of an underlying broadcast primitive, which is a best effort broadcast. 
This means that when a correct validator broadcasts a value, all the correct validators eventually deliver it. 
Messages are generated with a digital signature, providing a mechanism for cryptographic identification and validation in the protocol. 

We assume a \emph{partially synchronous model} \cite{dwork_consensus_1988}, where the system transitions from an asynchronous state to a synchronous state after an a-priori unknown Global Stabilization Time (\texttt{GST}).
Before \texttt{GST} there is an asynchronous period in which there is no bound on the message transfer delay $\Delta$. After \texttt{GST} there is a known finite bound on $\Delta$.
Note that all messages sent before \texttt{GST} are received at most at time $\texttt{GST}+\Delta$.
Note that even if we have synchronized clocks, having an asynchronous network before \texttt{GST} still makes the system partially synchronous.

For a significant portion of our analysis, we assume a network configuration wherein, during asynchronous periods, honest validators are separated into two distinct partitions. Communication between these partitions is restricted, reflecting a scenario where two regions are temporarily isolated yet maintain internal communication. This setup emulates the scenario where two regions of the world are temporarily unreachable from each other, while maintaining unaffected communication within each region.

\paragraph*{Fault Model}
Validators fall into two categories: \emph{honest} and \emph{Byzantine}. 
Honest validators, also called correct validators, adhere to the protocol, while Byzantine validators may arbitrarily deviate from the protocol\footnote{Since in this paper we are only interested in the consensus part of the protocol, we only characterize validator's behavior. For clients submitting transactions, as in any blockchain, we assume they can be Byzantine without impact on our analysis.}.
Following the literature \cite{castro_practical_1999}, we allow for a strong adversary that can coordinate Byzantine validators, even across network partitions\footnote{In our analysis, only the first three attacks leverage this power of the adversary because they occur during the asynchronous period, in which partition can occur. Meanwhile, the probabilistic bouncing attack happens in the synchronous period, so this power of the adversary is not necessary for the attack.}, thereby remaining unaffected by such partitions. But contrary to Castro and Liskov \cite{castro_practical_1999}, the adversary does not manipulate message delay between honest validators. 
We denote by $\beta_0$ the initial proportion of Byzantine validators' stake, with $\beta_0<1/3$.

The Ethereum Proof-of-Stake (PoS) protocol aspires to achieve Byzantine Fault Tolerance (BFT), ensuring the preservation of Safety and Liveness properties for any initial Byzantine stake proportion ($\beta_0$) strictly below 1/3.

\paragraph*{Ethereum PoS Properties} Validators keep a local data structure in form of a tree containing all the blocks perceived, then a consensus protocol helps to choose a unique chain in the tree. Ethereum has a particular trait that consists in having a \emph{finalized} chain as prefix of a chain vulnerable to forks.
A metaphor for this is that the finalized chain is the trunk that supports possibly various branches, and as time passes, the trunk grows and branches are trimmed\footnote{In the the paper, we use the terms ``chain'' and ``branch'' interchangeably.
}.

Intuitively, the Safety property of Ethereum states that the  finalized chain is not forkable, while the Liveness property states that the finalized chain always grows. The nuance with respect to classical consensus protocol is the existence of an Availability property on the entire chain that guarantees a constant growth of the chain despite failures and network partitions.  We report here a simplified version of the formal Ethereum PoS properties defined in \cite{pavloff_ethereum_2023} for enhanced clarity and self-contained reference, as follows:

\begin{definition}[\textbf{Candidate chain}]
 The candidate chain is the chain designated as the one to build upon according to the fork choice rule.
\end{definition}
The candidate chain can be seen as the entire chain, from the genesis to the last block perceived by a validator. The blocks in the candidate chain can be finalized or not.

\begin{definition}[\textbf{Finalized block}]
A block is finalized for a validator  if and only if  the block cannot be revoked, i.e., it permanently belongs to the validator's candidate chain.
\end{definition}

It stems from the definition that all the predecessors of a finalized block are also finalized.

\begin{definition}[\textbf{Finalized chain}]
The finalized chain is the chain constituted of all the finalized blocks.
\end{definition}

The finalized chain is always a prefix of any candidate chain.

Safety, Availability, and Liveness are expressed as follows:
\begin{property}[\textbf{Safety}]
A blockchain achieves \textbf{safety} if for any two correct validators with a finalized chain, then one chain  is necessarily the prefix of the other. 
\end{property}
\begin{property}[\textbf{Availability}]
A blockchain is \textbf{available} if the following two conditions hold: (1) any correct validator is able to append a block to its candidate chain in bounded time, regardless of the failures of other validators and the network partitions;  (2) the candidate chains of all correct validators are eventually growing.
\end{property}
\begin{property}[\textbf{Liveness}]
A blockchain is \textbf{live} if the finalized chain is eventually growing.
\end{property}

Accordingly to the aforementioned definitions, this paper considers forks within the finalized chain as a loss of Safety. As explained in the subsequent section, forks occurring within the candidate chain suffix, which has not yet been finalized, are resolved by the fork choice rule of the protocol.  This rule determines the chain upon which validators vote and build. 
However, this rule has not been explicitly designed to handle forks impacting the finalized chain. 
The protocol is not intended to fork the finalized chain, as the finalization process depends on a super-majority vote, ensuring Safety when the Byzantine stake is less than one third, i.e., $\beta_0<1/3$.
We look at two types of Safety loss: (1) the finalization of two conflicting chains, and (2) the break of the Safety threshold, meaning the Byzantine stake proportion of more than one-third.

\section{Ethereum protocol} \label{sec:ethProtocol}

In this work we study a specific mechanism of the Ethereum protocol, the \textit{inactivity leak}. To help our explanation and analysis of the inactivity leak, we first introduce some\footnote{We omit here a lot of details for the sake of space and simplicity. An interested reader must refer to \cite{ buterin_combining_2020, pavloff_ethereum_2023, neu_ebb_2021, edington_technical_2023}  for more thorough explanation of the protocol. } necessary understanding of the protocol. 

\subsection{Structure}
\label{subesc:structure}
In Ethereum PoS, time is measured in \emph{slots} and \emph{epochs}. 
A slot lasts 12 seconds.
Slots are assigned with consecutive numbers; the first slot is slot $0$. 
Slots are encapsulated in \emph{epochs}. An epoch is composed of 32 slots, thus lasting 6 minutes and 24 seconds. The first epoch (epoch 0) contains from slot 0 to slot 31; then epoch 1 contains slot 32 to 63, and so on. These slots and epochs allow associating the validators' roles to the corresponding time frame.
An essential feature of epochs is the \emph{checkpoint}. 
A checkpoint is a pair block-epoch $(b,e)$ where $b$ is the block of the first slot of epoch $e$.

An epoch can be seen as a round of a BFT Consensus in which all validators vote to reach consensus about which checkpoint to finalize. 

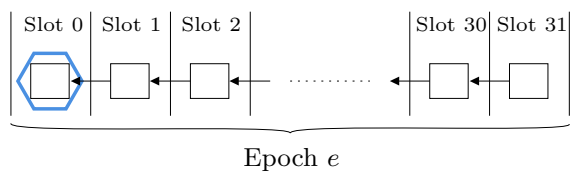
\begin{figure}[htbp]
    \centering
    \resizebox{.55\columnwidth}{!}{
    \begin{tikzpicture}[x=0.75pt,y=0.75pt,yscale=-1,xscale=1]

\draw   (87.05,80) -- (109.05,80) -- (109.05,100) -- (87.05,100) -- cycle ;
\draw   (132.69,80) -- (154.69,80) -- (154.69,100) -- (132.69,100) -- cycle ;
\draw    (132.69,90) -- (112.5,90) ;
\draw [shift={(109.5,90)}, rotate = 360] [fill={rgb, 255:red, 0; green, 0; blue, 0 }  ][line width=0.08]  [draw opacity=0] (5.36,-2.57) -- (0,0) -- (5.36,2.57) -- cycle    ;
\draw   (269.62,80) -- (291.62,80) -- (291.62,100) -- (269.62,100) -- cycle ;
\draw   (315.26,80) -- (337.26,80) -- (337.26,100) -- (315.26,100) -- cycle ;
\draw    (269.62,90) -- (249.8,90) ;
\draw [shift={(246.8,90)}, rotate = 360] [fill={rgb, 255:red, 0; green, 0; blue, 0 }  ][line width=0.08]  [draw opacity=0] (5.36,-2.57) -- (0,0) -- (5.36,2.57) -- cycle    ;
\draw    (315.26,90) -- (295.5,90) ;
\draw [shift={(292.5,90)}, rotate = 360] [fill={rgb, 255:red, 0; green, 0; blue, 0 }  ][line width=0.08]  [draw opacity=0] (5.36,-2.57) -- (0,0) -- (5.36,2.57) -- cycle    ;
\draw  [dash pattern={on 0.84pt off 2.51pt}]  (189.75,90) -- (235.39,90) ;
\draw    (121.28,50) -- (121.28,110) ;
\draw    (166.93,50) -- (166.93,110) ;
\draw    (258.21,50) -- (258.21,110) ;
\draw    (303.85,50) -- (303.85,110) ;
\draw    (349.49,50) -- (349.49,110) ;
\draw    (29.96,114.4) .. controls (28.56,121.11) and (189.33,114.28) .. (189.87,120) ;
\draw    (189.87,120) .. controls (189.36,113.71) and (350.89,120.31) .. (349.49,114.88) ;
\draw    (178.34,90) -- (158,90) ;
\draw [shift={(155,90)}, rotate = 360] [fill={rgb, 255:red, 0; green, 0; blue, 0 }  ][line width=0.08]  [draw opacity=0] (5.36,-2.57) -- (0,0) -- (5.36,2.57) -- cycle    ;
\draw  [color={rgb, 255:red, 74; green, 144; blue, 226 }  ,draw opacity=1 ][line width=1.5]  (70.64,90.02) -- (61.48,106.02) -- (43.16,106.02) -- (34,90.02) -- (43.16,74.03) -- (61.48,74.03) -- cycle ;
\draw    (87.05,90) -- (67.23,90) ;
\draw [shift={(64.23,90)}, rotate = 360] [fill={rgb, 255:red, 0; green, 0; blue, 0 }  ][line width=0.08]  [draw opacity=0] (5.36,-2.57) -- (0,0) -- (5.36,2.57) -- cycle    ;
\draw    (75.64,50) -- (75.64,110) ;
\draw    (30,50) -- (30,110) ;
\draw   (41.32,80.02) -- (63.32,80.02) -- (63.32,100.02) -- (41.32,100.02) -- cycle ;

\draw (54.5,57.14) node  [font=\scriptsize,xscale=1.25,yscale=1.25]  {$\mathrm{Slot} \ 0$};
\draw (190.6,135.55) node  [font=\small,xscale=1.25,yscale=1.25]  {$\mathrm{Epoch} \ e$};
\draw (99,57.14) node  [font=\scriptsize,xscale=1.25,yscale=1.25]  {$\mathrm{Slot} \ 1$};
\draw (144.65,57.14) node  [font=\scriptsize,xscale=1.25,yscale=1.25]  {$\mathrm{Slot} \ 2$};
\draw (282.11,57.14) node  [font=\scriptsize,xscale=1.25,yscale=1.25]  {$\mathrm{Slot} \ 30$};
\draw (328.32,57.14) node  [font=\scriptsize,xscale=1.25,yscale=1.25]  {${\textstyle \mathrm{Slot} \ 31}$};

\end{tikzpicture}
}
    \caption{Ethereum protocol Structure}
    \label{fig:slotEpoch}
\end{figure}

\subsection{Consensus}
\label{subesc:consensus}
Validators have two main roles: \emph{proposer} and \emph{attester}. The proposer's role consists in proposing a block during a specific slot;
this role is pseudo-randomly given to 32 validators by epoch (one for each slot). The attester's role consists in producing an attestation sharing the validator's view of the chain; this role is given once by epoch to each validator. 

The attestation contains two votes, a \textit{block vote} and a \textit{checkpoint vote}. 
The block vote is used in the \textit{fork choice rule} which determines the chain to vote and build upon for validators.
As its name suggests, the checkpoint vote points to checkpoints constituting the chain. 
It is
used to justify and finalize blocks to grow the finalized chain. 
Justification is the step prior to finalization.  
If validators controlling more than two-thirds of
the stake make the same checkpoint vote, then the checkpoint target is said justified. Finalization occurs when there are two consecutive justified checkpoints (one in epoch $e$ and the following one in epoch $e+1$).

Let us note that if justification occurs only every other epoch, finalization is not possible.

\subsection{Incentives}
The Ethereum PoS protocol provides validators with rewards and penalties to incentivize timely responses reaching consensus. 
There are three different types of penalties: slashing, attestation penalties, and inactivity penalties. (i) \emph{Slashing penalties.} Validators face slashing if they provably violate specific protocol rules, resulting in a partial loss of their stake and expulsion from the validator set.  (ii) \emph{Attestation penalties.} To incentivize timely and correct attestations (votes), the protocol rewards validators for adhering to the protocol and penalizes those who do not. If an attestation is missing or belatedly incorporated into the chain, its validator gets penalized.
(iii) \textbf{\emph{Inactivity penalties.}} Each epoch a validator is deemed inactive, its \emph{inactivity score} increments. However, if the protocol is not in an inactivity leak, all inactivity scores are reduced.
When finalization occurs regularly, a validator that is deemed inactive only gets attestation penalties. This changes when there is no finalization for four consecutive epochs: the inactivity leak begins. During the inactivity leak, that starts when there is no finalization for four consecutive epochs, all validators will get inactivity penalties directly linked to their stake and \emph{inactivity score}. The inactivity score varies with the validator's activity.
In addition to penalties, rewards are attributed for timely/correct attestation but not during the inactivity leak.
Our analysis of the impact of the inactivity leak on the protocol takes into consideration the slashing and inactivity penalties across 5 different scenarios (cf. \autoref{sec:analysis}).

Having provided a comprehensive overview of the Ethereum PoS consensus mechanism, we are now well-positioned to delve into the specifics of the \textit{inactivity leak}.

\section{Inactivity Leak}\label{sec:inactivityLeak}

The Ethereum PoS blockchain strives for the continuous growth of the finalized chain. In consequence, the protocol incentivizes validators to finalize blocks actively. In the absence of finalization, validators incur penalties.

The inactivity leak, introduced in \cite{buterin_casper_2017}, serves as a mechanism to regain finality. Specifically, if a chain has not undergone finalization for four consecutive epochs, the inactivity leak is initiated.
During the inactivity leak, the stakes of \textit{inactive} validators are drained until active validators amount for two-thirds of the stake.
A validator is labeled as inactive for a particular epoch if it fails to send an attestation or sends one with a wrong target checkpoint.

During the inactivity leak, there are no more rewards given to attesters\footnote{Actually, the only rewards to remain are for the block producers and the sync committees.}, and additional penalties are imposed on inactive validators.


\subsection{Inactivity Score}
The \emph{inactivity score} is a dynamic variable that adjusts based on a validator's activity. The inactivity score of a validator is determined based on the attestations contained in the chain. It is important to note that if there are multiple branches, a validator's inactivity score depends on the selected branch. Within an epoch, being active on one branch implies\footnote{This is true as long as the chain differ for at least one epoch.} inactivity on another (for honest validators).

More precisely, the inactivity score is updated every epoch: if validator $i$ is active, then its inactivity score is reduced by 1; otherwise, 4 is added to it. When the inactivity leak is not in place, every epoch the inactivity scores are decreased by 16, which often nullifies low inactivity scores. 

During an inactivity leak, at epoch $t$, the inactivity score, $I_i(t)$, of validator $i$ is:
\begin{equation}
    \begin{cases}
        I_i(t) = I_i(t-1)+4, \text{if $i$ is inactive at epoch $t$} \\
        I_i(t) = \max(I_i(t-1)-1, 0), \text{ otherwise.}
    \end{cases}
\end{equation}
Each attester has thus an inactivity score that fluctuates depending on its (in)activity. In the protocol, the inactivity score is always superior to zero.
A validator's inactivity for epoch $t$ is determined by whether it sent an attestation for this epoch or if the sent attestation contains a correct checkpoint vote. Here ``correct'' implies that the target of the checkpoint vote belongs to the chain considering it.


\subsection{Inactivity penalties}
Validators that are deemed inactive incur penalties.
Let $s_i(t)$ represent the stake of validator $i$ at epoch $t$, and let $I_i(t)$ denote its inactivity score. 
The penalty at each epoch $t$ is $I_i(t-1)\cdot s_i(t-1)/2^{26}$. 
Therefore, the evolution of the stake is expressed by:
\begin{equation}\label{eq:discreteStake}
s_i(t)=s_i(t-1)-\frac{I_i(t-1)\cdot s_i(t-1)}{2^{26}}.    
\end{equation}


\subsection{Stake's functions during an inactivity leak}\label{subsec:stakeFunctions}

In this work, we model the stake function $s$ (see \autoref{eq:discreteStake}) as a continuous and differentiable function, yielding the following differential equation:
\begin{equation}\label{eq:stakeFunction}
s'(t)=-I(t)\cdot s(t)/2^{26}.
\end{equation}

We then explore three distinct validator's behaviors during inactivity leak, each influencing their inactivity score, and consequently, their stake.

\begin{enumerate}[(a)] 
    \item Active validators: they are always active.
    \item Semi-active validators: they are active every two epochs.
    \item Inactive validators: they are always inactive.
\end{enumerate}

Note that, in case of a fork, this categorization depends on the specific branch under consideration as different branches may yield different evaluation of each validator's behavior. 

Validators with these three behaviors experience a different evolution in their inactivity score: (a) Active validators have a constant inactivity score $I(t)=0$; (b) Semi-active validators' inactivity score increases by 3 every two epochs ($+4-1$), their inactivity score on average at $t$ is $I(t)=3t/2$; (c) Inactive validators' inactivity score increases by 4 every epoch $I(t)=4t$. Thanks to \autoref{eq:stakeFunction}, we can determine the evolution of the stake of each type of validator during an inactivity leak:
\begin{enumerate}[(a)] 
    \item Active validator's stake: $ s(t) = s_0 = 32. $
    \item Semi-active validator's stake: $s(t)=s_0e^{-3t^2/2^{28}}.$
    \item Inactive validator's stake: $ s(t) = s_0e^{-t^2/2^{25}}. $
\end{enumerate}

This categorization is orthogonal to the Byzantine-Honest categorization.
For instance, an honest validator can appear inactive in one branch due to poor connectivity or an asynchronous period (due to network partition or congestion). On the other hand, a Byzantine validator intentionally chooses one of these behaviors (e.g., being semi-active) to execute the attacks.

\begin{figure}[htbp]
    \centerline{\includegraphics[scale=0.55]{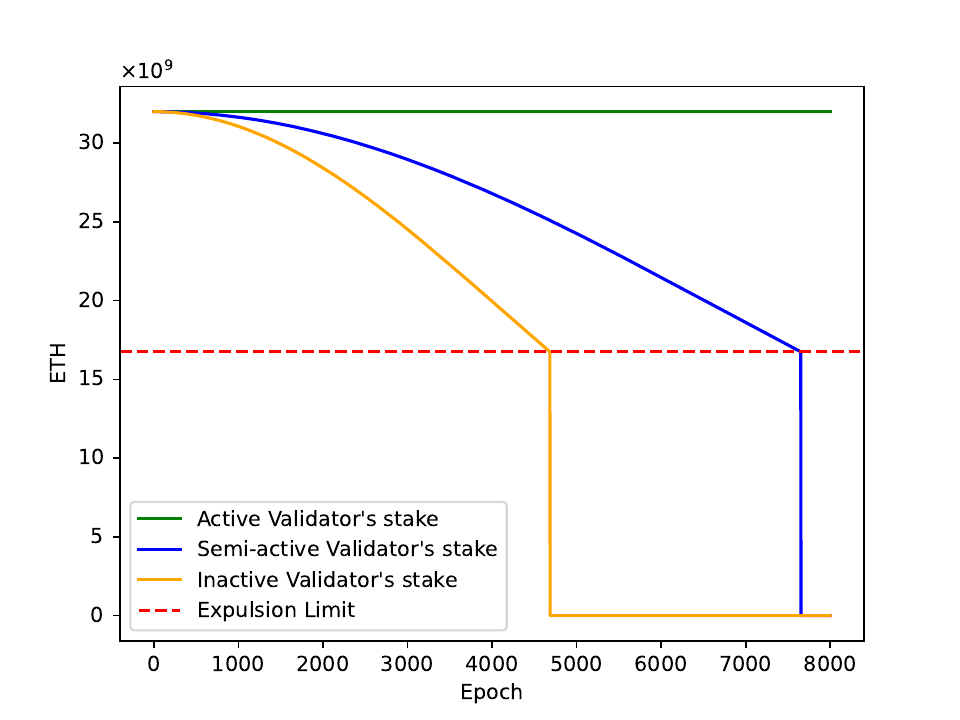}}
    \caption{This figure shows 3 different stake's trajectories in the event of an inactivity leak: the stake of a validator active every epoch, the stake of a validator active every other epoch (semi-active) and an inactive validator. The inactive validators get ejected at epoch $t=4685$. The semi-active validators get ejected at $t=7652$. For reference, 5000 epochs is about 3 weeks.}
    \label{fig:validatorsStake}
\end{figure}

We illustrate in \autoref{fig:validatorsStake} the evolution of the validators' stake depending on their behaviors. We also account for the ejection of validators with a stake lower or equal than 16.75.

Using these newly defined stake functions, we explore five different scenarios in \autoref{sec:analysis}.

The first scenario, with only honest validators, serves as a baseline to assess the impact of Byzantine validators. Even in this seemingly straightforward setting, Safety is compromised.

In the second scenario, Byzantine validators come into play and aim at expediting finalization on conflicting branches. They do so by performing slashable actions. Thus, they will get ejected from the set of validators once communication is restored among honest validators and evidence of their slashable offense is included in a block. 
We outline their impact based on their initial stake proportion. With an initial stake proportion of $\beta_0=0.2$, the finalization on conflicting chains occurs after 3107 epochs. With $\beta_0=0.33$, the conflicting finalization occurs only after 503 epochs.

In the third and forth scenarios Byzantine validators show non-slashable behaviors. Indeed, Byzantine validators are semi-active, which means they are active on both chains but in a non-slashable manner. However, in the third scenario they aim at finalizing on conflicting branches as soon as possible -- conflicting finalization in 556 epochs, for an initial stake proportion of $\beta_0=0.33$ -- while in the fourth scenario they aim for their stake proportion to go over the 1/3 threshold.

The last scenario delves into the effect of the probabilistic bouncing attack regarding the Byzantine stake proportion taking the inactivity leak into account. In this attack Byzantine validators initially aim at delaying finality by being alternatively active (bounce) on both chains of a fork. This confuses honest validators that also bounce from one chain to the other.
We detail how to find the distribution of honest validators' stake in this setting considering the inactivity penalties. We also cover how the Byzantine validators' stake proportion can go over 1/3 if their initial proportion is close enough to 1/3.

The scenarios unfold within the context of a partially synchronous network while offering a meticulous examination of the property of Safety and the evolution of the proportion of Byzantine validators. 
The variations in initial conditions and outcomes of each scenario are summarized in \autoref{tab:recapScenario}. 
\begin{table}
\caption{Analysed scenarios associated with their outcomes. Initially the proportion of Byzantine's stake is smaller than 1/3 and is zero for the first scenario. }
\begin{center}
\begin{tabular}{|c|c|c||c|}
\hline
Scenario  & Outcomes \\ \hline \hline
\ref{subsec:HonDoubleFinalization} \hfill All honest \hfill\null &  2 finalized branches \\ 
\ref{subsubsec:withSlashing}  \hfill Slashable Byzantine \hfill\null & 2 finalized branches \\
\ref{subsubsec:withoutSlashing}  \hfill Non slashable Byzantine \hfill\null & 2 finalized branches \\
\ref{subsubsec:oneThirdByz}  \hfill Non slashable Byzantine \hfill\null & $\beta>1/3$ \\
\ref{subsec:revisitPBA}  \hfill Probabilistic Bouncing attack \hfill\null & $\beta>1/3$ probably \\ \hline
\end{tabular}
\end{center}
\label{tab:recapScenario}
\end{table}

\section{Analysis}\label{sec:analysis}

In this section we study the robustness of the Safety property within the context of the inactivity leak. By construction, in case of a prolonged partition, two different chains can potentially be finalized, leading to conflicting finalized blocks. We delineate scenarios that can produce such predicament.

Considering the presence of Byzantine validators, we study how the proportion of Byzantine validators' stake evolves during an inactivity leak.
Furthermore, we are interested in scenarios where the inactivity leak mechanism becomes the backbone of an attack strategy, potentially causing the proportion of Byzantine stakes to exceed the 1/3 security threshold  (cf. \autoref{subsubsec:oneThirdByz} and \autoref{subsec:revisitPBA}).

\subsection{GST upper bound for Safety}
\label{subsec:HonDoubleFinalization}
In this first subsection, we look for an upper bound on \texttt{GST} before which no finalization on conflicting chains can happen in case of a partition. 
We study the case of an inactivity leak with these conditions: (i) Only honest validators, no Byzantine validators, and (ii) the network is asynchronous (before \texttt{GST}).

In case of catastrophic events, during an instance of a particularly disrupted network, an arbitrary large set of honest validators might be unreachable before \texttt{GST}. During this asynchronous period, the subset of validators still communicating with each other will continue to try to finalize new blocks. We assume that, within each partition, the message delay is bounded as in the synchronous period; however, communication between partitions is not restored before the GST. As mentioned in the system model, Byzantine validators can communicate between partitions without restriction but cannot manipulate the message delay between honest validators. The active validators must represent more than two-third of the stake to be able to finalize.
After 4 epochs without finalization, the inactivity leak starts.

All the validators deemed inactive will get their stake reduced progressively. This will continue until the active validator's constitute at least two thirds of the stake and finalize anew.

\

\paragraph{Two finalized chains} A noteworthy scenario arises during asynchronous periods that can lead to a network partition and the creation of two distinct finalized chains. 
If this partition persists for an extended period, both chains independently drain the stakes of validators they consider inactive until they finalize again. 
Although the protocol permits this behavior by design, it results in the finalization of two conflicting chains, thereby compromising the Safety property.

This outcome is in line with Ethereum PoS prioritizing Liveness of Safety. But to the best of our knowledge, this corner case has not been discussed in details. 

\

We can theoretically assess the time required to finalize both branches of the fork. Suppose honest validators remain on their respective branches due to the partition. In this case, by understanding the distribution of these validators across the partitions, we can compute the time it takes for the proportion of active validators' stake to return to 2/3 of the stake on each branch, permitting new finalization.

Let $n_{\rm H}$ and $n_{\rm B}$ denote the initial number of honest validators and Byzantine validators at the beginning of the inactivity leak ($n_{\rm H}+n_{\rm B}=n$). Additionally, $n_{\rm H_1}$ and $n_{\rm H_2}$ represents the number of honest validators active on branch 1 and on branch 2, respectively ($n_{\rm H_1}+n_{\rm H_2}=n_{\rm H})$.

We denote by $p_0=n_{\rm H_1}/n_{\rm H}$ the initial proportion of honest validators remaining active on branch $1$, and $1-p_0=n_{\rm H_2}/n_{\rm H}$ represents the proportion of honest validators active on branch $2$ (hence inactive on branch $1$). 
In this first scenario, with only honest validators and no Byzantine validators, $p_0$ represents the proportion of all validators active on branch 1. Indeed, since $n_{\rm H}/n=1$ we have that $n_{\rm H_1}/n_{\rm H}\times n_{\rm H}/n=p_0$.

We have assessed how validators' stakes vary based on their level of activity, we can in consequence express the ratio of active validators on branch 1 at time $t$:
\begin{equation}\label{eq:preliHonestActiveRatio}
\frac{n_{\rm H_1}s_{\rm H_1}(t)}{n_{\rm H_1}s_{\rm H_1}(t)+n_{\rm H_2}s_{\rm H_2}(t)} ,
\end{equation}
with $s_{\rm H_1}$ and $s_{\rm H_2}$ being the stake of honest active and inactive validators, respectively. We know the function of their stake according to time, and by dividing in the numerator and the denominator by the total number of validators ($n=n_{\rm H}$), we can rewrite \autoref{eq:preliHonestActiveRatio} as:
\begin{equation}\label{eq:honestActiveRatio}
\frac{p_0}{p_0+(1-p_0)e^{-t^2/2^{25}}}.    
\end{equation}
The initial stake value $s_0$ is factored out of the equation.
This function is critical as the moment it reaches 2/3 or more, finalization can occur\footnote{With a proportion of two-thirds of validators' stake active justification then finalization can occur in 2 epochs.} on the branch.

\begin{figure}[htbp]
    \centerline{\includegraphics[scale=0.55]{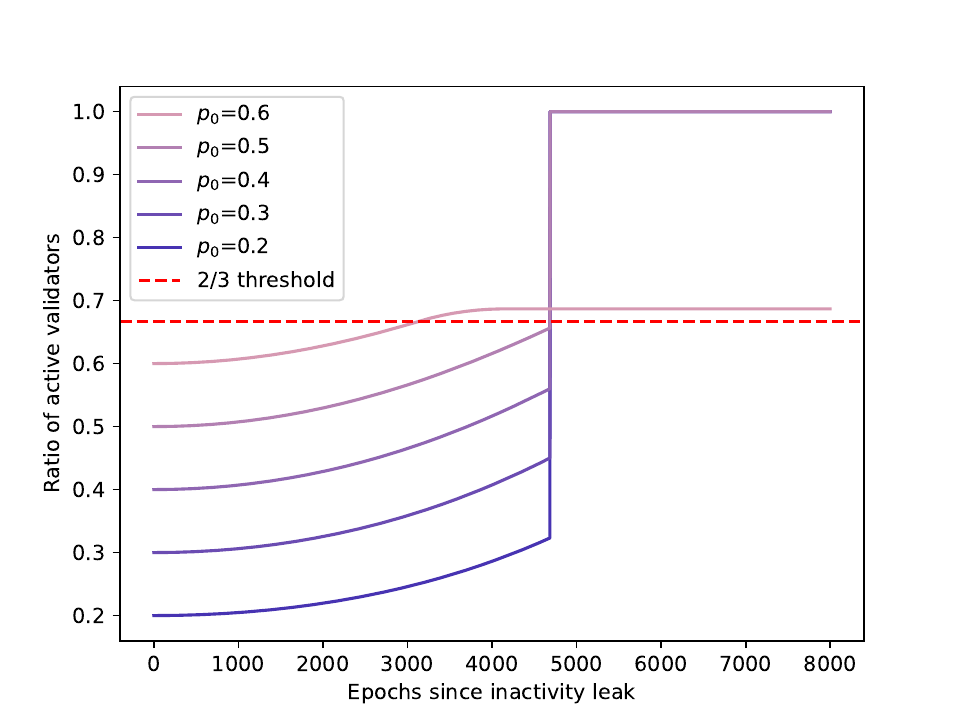}}
    \caption{Evolution of the ratio of active validators depending on the proportion $p_0$ of active validators on the branch. This follows the ratio given in \autoref{eq:honestActiveRatio} before regaining 2/3 of active validators or 
    the expulsion of inactive validators at epoch $t=4685$.}
    \label{fig:honestActiveRatio}
\end{figure}

To establish the upper bound on \texttt{GST} under which two conflicting branches finalize, we must find when finalization occurs on each branch, for each initial proportion of active validators $p_0$ and inactive validators $1-p_0$. 
We simulate the evolution of the ratio of active validators (\autoref{eq:honestActiveRatio}) during an inactivity leak with different values of $p_0$ in \autoref{fig:honestActiveRatio}. The simulation starts with both active and inactive validators at 32 ETH. At epoch 0, the inactivity leak begins.

For $p_0=0.5$ or less, the ratio jumps to 1 at $t=4685$, this is due to the fact that validators with a stake below 16.75 ETH are ejected from the set of validators. 
Conversely for $p_0=0.6$, the proportion of active validators do not jump drastically as 2/3 of active validators is regained before the ejection of inactive validators, permitting the active validator to finalize, hence ending the inactivity leak. Interestingly, with $p=0.6$ we can see that the ratio still increases several epochs after the proportion of 2/3 of active validators' stake is reached. This is because the penalties for inactive validators take some time to return to zero.

As expected and shown by \autoref{fig:honestActiveRatio}, a chain with more active validators will regain finality quicker. To ascertain how quick, we seek when the ratio is equal to 2/3. Taking into account the expulsion\footnote{We drew inspiration for this initial work from the insights presented in \cite{edington_technical_2023}.} of inactive validators at $t=4685$, we can find the value $t$ at which the 2/3 threshold is reached:
\begin{equation}
      t = \min \left( \sqrt{2^{25}[\log(2(1-p_0))-\log(p_0)]}\; , 4685 \right).
\end{equation}

This calculation pertains to $0<p_0<2/3$ (when there is less that 2/3 of active validators) ensuring that the epoch $t$ can be computed.

Conflicting finalization occurs once the slowest branch to finalize has regained finality. 
Our observation highlights that the lower the proportion of active validators, the slower the branch will regain finality.
Hence, the fastest way to reach finality on both chains would be for honest validators to be evenly proportionate with half validators active on one chain and the other half on the other chain ($p_0=1-p_0=0.5$). 
In this case, the ratio of active validators amount to 2/3 on both chain at $t=4685$ epochs (about 3 weeks). We can note here that even with the best configuration to finalize quickly on conflicting branches, it is impossible to be faster than 4685 epochs. Thus with only honest validators, whatever their proportion on each branch, the last chain to finalize will always finalize at $t=4685$. 

\
 
Finality on both chains is achieved precisely at $4686$ epochs after the beginning of the inactivity leak. Adding an epoch is necessary after gaining 2/3 of active stake to finalize the preceding justified checkpoint. This finalization ends the inactivity leak which has lasted approximately 3 weeks. \textit{ Any network partition lasting longer than 4686 epochs will result in a loss of Safety because of conflicting finalization. This is an upper bound for Safety on the duration of the inactivity leak with only honest validators.}

\subsection{Upper bound decrease due to Byzantine validators}\label{subsec:ByzDoubleFinalization}

In a trivial setup with only honest validators, Safety does not hold if the inactivity leak is not resolved quickly. This prompts us to study the scenario in the presence of Byzantine validators to evaluate how much they will be able to hasten the conflicting finalization. 
We describe two possible outcomes, the first one violates Safety but Byzantine validators gets slashed, the second one violates Safety as well but no validators gets slashed. 
A slashing penalty entails an ejection of the validator set as well as a loss of part of the validator's stake. 
Both scenarios expedite the time $t$ at which Safety is breached, with different velocity depending on the chosen method.

We study the inactivity leak with these conditions: (i) at the beginning, less than one third of the stake is held by Byzantine validators ($\beta_0=n_{\rm B}/n<1/3$), the rest is held by honest validators ($1-\beta_0=n_{\rm H}/n$); (ii) The network is asynchronous (before \texttt{GST}); and (iii) Byzantine validators are not affected by network partitions.\footnote{
In a model without partitions, one needs to give Byzantine validators more power to recreate our scenario. They must be able to control the network delay to allow them to be active on both branches while preventing honest validators from even observing the branch on which they are not active. They can manipulate message delays between groups of honest validators to simulate a partition between them. }

The situation is the following:
\begin{itemize}
    \item Honest validators are divided into branches $1$ and $2$; a proportion $p_0=n_{\rm H_1}/n_{\rm H}$ of the honest validators are active on branch $1$ while a proportion $1-p_0=n_{\rm H_2}/n_{\rm H}$ are active on branch $2$. (Meaning that on branch $1$, a proportion $n_{\rm H_1}/n_{\rm H}\times n_{\rm H}/n=p_0(1-\beta_0)$ are honest and active and a proportion $n_{\rm H_2}/n_{\rm H}\times n_{\rm H}/n=(1-p_0)(1-\beta_0)$ are honest and inactive).
    \item Byzantine validators are not restricted to either partition, they are connected to both.
\end{itemize}

\subsubsection{With slashing}\label{subsubsec:withSlashing}

\begin{figure}[htbp]
    \centering
    \resizebox{.4\columnwidth}{!}{
    \begin{tikzpicture}[x=0.75pt,y=0.75pt,yscale=-1,xscale=1]

\draw    (50,125) -- (100,100) ;
\draw    (50,125) -- (100,150) ;
\draw  [color={rgb, 255:red, 255; green, 0; blue, 0 }  ,draw opacity=1 ][fill={rgb, 255:red, 255; green, 52; blue, 52 }  ,fill opacity=0.86 ] (100,100) -- (300.8,100) -- (300.8,110) -- (100,110) -- cycle ;
\draw  [color={rgb, 255:red, 255; green, 0; blue, 0 }  ,draw opacity=1 ][fill={rgb, 255:red, 255; green, 52; blue, 52 }  ,fill opacity=0.86 ] (100,150) -- (300.8,150) -- (300.8,160) -- (100,160) -- cycle ;
\draw    (100,100) -- (300.8,100) ;
\draw    (100,150) -- (300.8,150) ;
\draw  [dash pattern={on 0.84pt off 2.51pt}]  (100,60) -- (100,170) ;
\draw  [dash pattern={on 0.84pt off 2.51pt}]  (150,60) -- (150,170) ;
\draw  [dash pattern={on 0.84pt off 2.51pt}]  (200,60) -- (200,170) ;
\draw  [dash pattern={on 0.84pt off 2.51pt}]  (250,60) -- (250,170) ;

\draw (128,70) node  [font=\small,xscale=1.25,yscale=1.25]  {$t_{1}$};
\draw (178,70) node  [font=\small,xscale=1.25,yscale=1.25]  {$t_{2}$};
\draw (228,70) node  [font=\small,xscale=1.25,yscale=1.25]  {$t_{3}$};
\draw (272,70) node  [font=\small,xscale=1.25,yscale=1.25]  {$t_{4}$};
\draw (78,70) node  [font=\small,xscale=1.25,yscale=1.25]  {$t_{0}$};

\end{tikzpicture}
    }
    \caption{Byzantine validators active on both chains of a fork at the same time during asynchronous times.}
    \label{fig:ByzSlashingSchema}
\end{figure}
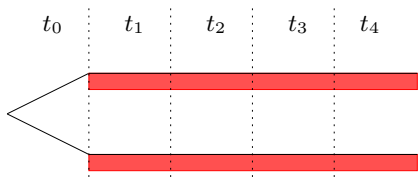

In the event of a fork during asynchronous times, Byzantine validators can be active on the two branches (\autoref{fig:ByzSlashingSchema}). Being active on two branches means sending correct attestations on both every epoch.
Such behavior is considered a slashable offense, incurring penalties, but only if detected by honest validators. The slashable offense is punished once a proof of conflicting attestation during the same epoch has been included in a block.
Thus, before \texttt{GST}, Byzantine validators could operate on both branches without facing punishment as long as honest validators are unaware of the conflicting attestations. Byzantine validators have control over the message delay before \texttt{GST} rending this behavior possible. They can thereby expedite the finalization on different branches. 

We study here the time needed for finalization to occur on conflicting branches depending on the proportion of Byzantine validators. 
The ratio of active validators at epoch $t$ is: 
\begin{equation}
    \frac{n_{\rm H_1}s_{\rm H_1}(t)+n_{\rm B}s_{\rm B}(t)}{n_{\rm H_1}s_{\rm H_1}(t)+n_{\rm B}s_{\rm B}(t)+n_{\rm H_2}s_{\rm H_2}(t)},
\end{equation}
with $s_{\rm H_1}$, $s_{\rm B}$ and $s_{\rm H_2}$ being the stake of honest active, Byzantine active and honest inactive validators, respectively. This can be rewritten as:
\begin{equation}
     \frac{p_0(1-\beta_0)+\beta_0}{p_0(1-\beta_0)+\beta_0+(1-p_0)(1-\beta_0)e^{-t^2/2^{25}}},
\end{equation}
where $\beta_0$ represents the initial proportion of Byzantine validators, and $p_0$ denotes the initial proportion of honest active validators. In contrast to the analysis with only honest validators (cf. \autoref{eq:honestActiveRatio}), here, Byzantine validators are present and active on both chains. 
Nonetheless, as before, we can obtain the ratio of active validators on the other branch just by interchanging $p_0$ and $1-p_0$. 
Finality on conflicting branches occurs when the last of the two branches finalizes. Similarly to the previous example, the branch with the fewer initial honest active validators ($p_0$) will finalize the latest. This happens $t$ epochs after the beginning of the inactivity leak, with
\begin{equation}\label{eq:byzTimeForFinalization}
      t = \min\left(\sqrt{2^{25}\left[\log(2(1-p_0))-\log(p_0+\frac{\beta_0}{1-\beta_0})\right]}, 4685 \right).
\end{equation}

Finality on conflicting branches is achieved the quickest when honest validators are evenly split between the branches of the fork, for $p_0=0.5$. 


\begin{table}[htbp]
\caption{Time before finalization on conflicting branches depending on the initial proportion of Byzantine validators $\beta_0$ for $p_0=0.5$ with slashing behaviour based on \autoref{eq:byzTimeForFinalization}.}
\begin{center}
\begin{tabular}{|c|c|}
\hline
\textbf{$\beta_0$} & \textbf{$t$} \\
\hline
\textbf{0} & \textbf{4685} \\
 0.1 & 4066 \\ 
 0.15 & 3622 \\ 
 0.2 & 3107 \\
 0.33 & 502 \\
\hline
\end{tabular}
\label{tab:finalizationSlashing}
\end{center}
\end{table}
\autoref{tab:finalizationSlashing} gives the epoch at which concurrent finalization occurs for $p_0=0.5$. This outline the rapidity at which finality can be regained depending on the initial proportion $\beta_0$ of Byzantine validators' stake. The table shows that 503 epochs (approximately 2 days) could suffice to finalize blocks on two different chains, but hypothetically it could be quicker than that. 
In fact, as $\beta_0$ gets closer to 1/3, the number of epochs required before concurrent finalization occurs (\autoref{eq:byzTimeForFinalization}) approaches 0.
The explanation is that if $\beta_0$ were to start at exactly 1/3, then with $p_0=0.5$ it would mean that on each branch we would start with $p_0(1-\beta_0)+\beta_0=2/3$ of active validators, hence finalizing immediately. This explains why if $\beta_0$ is very close to 1/3, the proportion of active validators reaches 2/3 rapidly.
\textit{Hence, Byzantine validators can expedite the loss of Safety. If their initial proportion is 0.33, they can make conflicting finalization occur approximately ten times faster than scenarios involving only honest participants.}

\

One can notice that if Byzantine validators act in a slashable manner, they will be penalized after the asynchronous period ends. However, the harm is already done. Once the finalization on two branches has occurred, the branches are irreconcilable with the current protocol. Next, we demonstrate that Byzantine validators can employ more subtle strategies to break Safety without slashable actions. 

\

\subsubsection{Without slashing} \label{subsubsec:withoutSlashing}

\begin{figure}[htbp]
    \centering
    \resizebox{.7\columnwidth}{!}{
    \begin{tikzpicture}[x=0.75pt,y=0.75pt,yscale=-1,xscale=1,scale=0.6, every node/.style={scale=0.8}]

\draw    (50,125) -- (100,100) ;
\draw    (50,125) -- (100,150) ;
\draw  [color={rgb, 255:red, 255; green, 0; blue, 0 }  ,draw opacity=1 ][fill={rgb, 255:red, 255; green, 52; blue, 52 }  ,fill opacity=0.86 ] (100,100) -- (150,100) -- (150,110) -- (100,110) -- cycle ;
\draw  [color={rgb, 255:red, 255; green, 0; blue, 0 }  ,draw opacity=1 ][fill={rgb, 255:red, 255; green, 52; blue, 52 }  ,fill opacity=0.86 ] (150,150) -- (200,150) -- (200,160) -- (150,160) -- cycle ;
\draw  [color={rgb, 255:red, 255; green, 0; blue, 0 }  ,draw opacity=1 ][fill={rgb, 255:red, 255; green, 52; blue, 52 }  ,fill opacity=0.86 ] (200,100) -- (250,100) -- (250,110) -- (200,110) -- cycle ;
\draw  [color={rgb, 255:red, 255; green, 0; blue, 0 }  ,draw opacity=1 ][fill={rgb, 255:red, 255; green, 52; blue, 52 }  ,fill opacity=0.86 ] (250,150) -- (300,150) -- (300,160) -- (250,160) -- cycle ;
\draw  [dash pattern={on 0.84pt off 2.51pt}]  (100,60) -- (100,170) ;
\draw  [dash pattern={on 0.84pt off 2.51pt}]  (150,60) -- (150,170) ;
\draw  [dash pattern={on 0.84pt off 2.51pt}]  (200,60) -- (200,170) ;
\draw  [dash pattern={on 0.84pt off 2.51pt}]  (250,60) -- (250,170) ;
\draw  [color={rgb, 255:red, 255; green, 0; blue, 0 }  ,draw opacity=1 ][fill={rgb, 255:red, 255; green, 52; blue, 52 }  ,fill opacity=0.86 ] (351,100) -- (451,100) -- (451,110) -- (351,110) -- cycle ;
\draw  [color={rgb, 255:red, 255; green, 0; blue, 0 }  ,draw opacity=1 ][fill={rgb, 255:red, 255; green, 52; blue, 52 }  ,fill opacity=0.86 ] (450,150) -- (550,150) -- (550,160) -- (450,160) -- cycle ;
\draw    (100,100) -- (310,100) ;
\draw  [dash pattern={on 0.84pt off 2.51pt}]  (350,60) -- (350,170) ;
\draw  [dash pattern={on 0.84pt off 2.51pt}]  (400,60) -- (400,170) ;
\draw  [dash pattern={on 0.84pt off 2.51pt}]  (450,60) -- (450,170) ;
\draw  [dash pattern={on 0.84pt off 2.51pt}]  (500,60) -- (500,170) ;
\draw  [dash pattern={on 0.84pt off 2.51pt}]  (300,60) -- (300,170) ;
\draw    (340,100) -- (550,100) ;
\draw    (100,150) -- (310,150) ;
\draw    (340,150) -- (550,150) ;
\draw  [dash pattern={on 4.5pt off 4.5pt}]  (310,100) -- (340,100) ;
\draw  [dash pattern={on 4.5pt off 4.5pt}]  (310,150) -- (340,150) ;

\draw (128,70) node  [font=\small,xscale=1.25,yscale=1.25]  {$t_{1}$};
\draw (178,70) node  [font=\small,xscale=1.25,yscale=1.25]  {$t_{2}$};
\draw (228,70) node  [font=\small,xscale=1.25,yscale=1.25]  {$t_{3}$};
\draw (272,70) node  [font=\small,xscale=1.25,yscale=1.25]  {$t_{4}$};
\draw (78,70) node  [font=\small,xscale=1.25,yscale=1.25]  {$t_{0}$};
\draw (378,70) node  [font=\small,xscale=1.25,yscale=1.25]  {$t_{n}$};
\draw (428,70) node  [font=\small,xscale=1.25,yscale=1.25]  {$t_{n+1}$};
\draw (478,70) node  [font=\small,xscale=1.25,yscale=1.25]  {$t_{n+2}$};
\draw (522,70) node  [font=\small,xscale=1.25,yscale=1.25]  {$t_{n+3}$};
\draw (329.5,70) node  [font=\small,xscale=1.25,yscale=1.25]  {$.\ .\ .$};

\end{tikzpicture}
}
    \caption{Byzantine validators active on both branches of a fork alternatively during asynchronous times.}
    \label{fig:ByzNoSlashingSchema}
\end{figure}
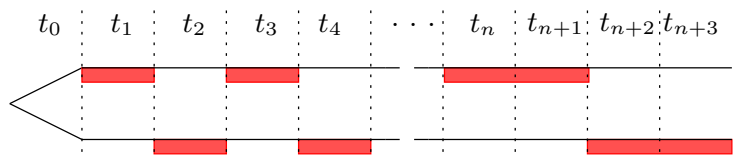

The Byzantine validators have a way to hasten the violation of the Safety property without incurring slashable offense. 
While not as rapid as being active on both branches simultaneously, they can be semi-active on both branches alternatively. Being semi-active on each branch means they are only active every other epoch.  
This approach diminishes their stake on each branch due to inactivity penalties. Nevertheless, at some point they will be able to finalize on two conflicting branches by being active two epochs in a row on one branch then on the other (see \autoref{fig:ByzNoSlashingSchema}). Byzantine validators will be able to finalize when the proportion of their stake plus the proportion of the stake of honest active validators is above 2/3 on the branch (cf. \autoref{eq:ByzSemiActiveRatio}).

At that point, Byzantine validators must remain active for two consecutive epochs on each branch to finalized them both.
If they are only semi-active, they can alternate justifications for checkpoints on each branch but will not achieve finalization. However, by maintaining activity for two consecutive epochs, first on one branch and then on the other, they ensure two sequential justifications, leading to the finalization of a checkpoint.

\

We gave the different evolution of stakes depending on the activity of validators (\autoref{subsec:stakeFunctions}). Now that Byzantine validators are semi-active, their stake follows the curve $s_0e^{-3t^2/2^{28}}$.
We simplify the ratio as previously and we get that finalization occurs on the branch when the ratio
\begin{equation}\label{eq:ByzSemiActiveRatio}
    \frac{p_0(1-\beta_0)+\beta_0 e^{-3t^2/2^{28}}}{p_0(1-\beta_0)+\beta_0 e^{-3t^2/2^{28}}+(1-p_0)(1-\beta_0)e^{-t^2/2^{25}}}
\end{equation}
goes over 2/3. With $\beta_0$ and $p_0$ being the initial proportion of Byzantine validators and the proportion of honest active validators on the branch, respectively. 

\begin{figure}
    \centering
    \includegraphics[scale=0.55]{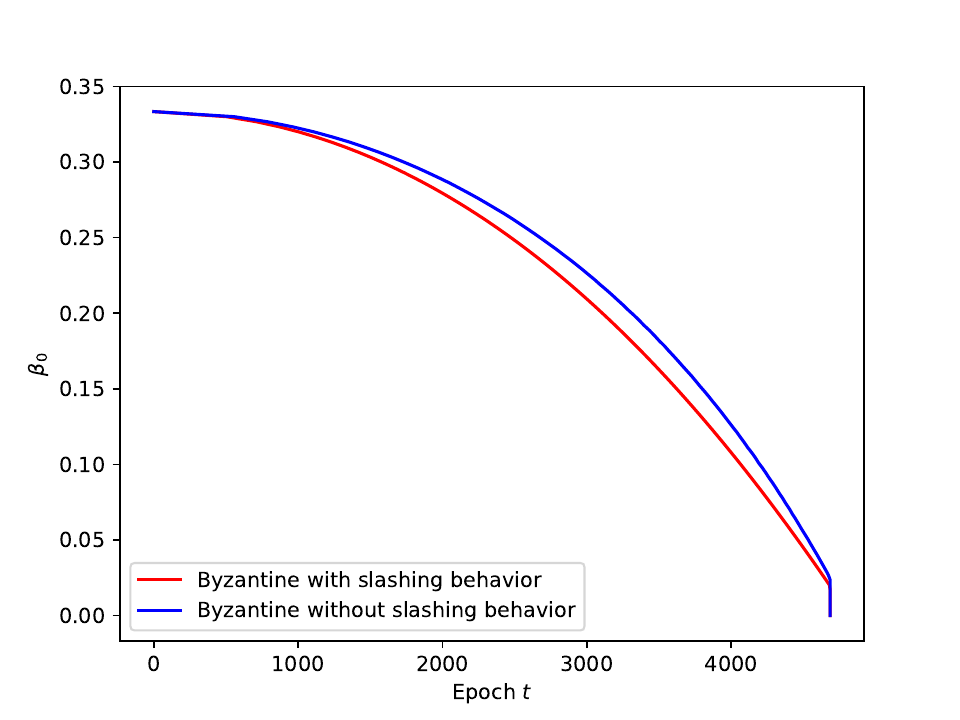}
    \caption{Time before finalization on conflicting branches, depending on the initial proportion of Byzantine validators $\beta_0$ and whether they engage in slashable actions.}
    \label{fig:byzSlashANDbyzNoSlash}
\end{figure}

In contrast to the previous scenario, obtaining an analytic solution for $t$ to determine the epoch when the ratio hits 2/3 is not straightforward. Therefore, we apply numerical methods on \autoref{eq:ByzSemiActiveRatio} with initial parameters $p_0=0.5$ and $\beta_0=0.33$, resulting in a calculated $t$ value of 555.65.
Meaning, it will take 556 epochs to finalize, about 2 days and a half. 
As previously, the proximity of $\beta_0$ to 1/3 significantly influences the speed of finalization, as outlined in \autoref{tab:finalizationNoSlashing} and \autoref{fig:byzSlashANDbyzNoSlash}. 
\autoref{fig:byzSlashANDbyzNoSlash} shows how the proportion of Byzantine affects the time of conflicting finalization. Let us notice that although the acceleration is not as pronounced as in the previous scenario, it remains noteworthy that Byzantine validators still exert a substantial impact on breaching Safety, while not committing any slashable offense.
\textit{Hence, Byzantine validators can expedite the loss of Safety without committing any slashable action. If their initial proportion is 0.33, they can make conflicting finalization occur approximately eight times faster than scenarios involving only honest participants.}

\begin{table}[htbp]
\caption{Time before finalization on conflicting branches depending on the initial proportion of Byzantine validators $\beta_0$ for $p_0=0.5$  without slashing behaviour based on \autoref{eq:ByzSemiActiveRatio}.}
\begin{center}
\begin{tabular}{|c|c|}
\hline
\textbf{$\beta_0$} & \textbf{$t$} \\
\hline
\textbf{0} & \textbf{4685} \\
 0.1 & 4221 \\
 0.15 & 3819\\
 0.2 & 3328\\
 0.33 & 556 \\
\hline
\end{tabular}
\label{tab:finalizationNoSlashing}
\end{center}
\end{table}

Another consequence of being ``semi-active'' on both branches is that Byzantine validators can decide when finalizing on each branch. Indeed, even when the proportion of their stake plus the proportion of honest and active validators' stake is above 2/3, finalization only occurs when the Byzantine validators stay active two consecutive epochs on the same chain. Being active for two epochs will justify the two consecutive epochs, thus finalizing an epoch.

There exists a scenario in which the Byzantine validators might delay finalization intentionally, aiming to increase their stake's proportion beyond the threshold of 1/3 without incurring slashing afterwards. 

\

\subsubsection{More than one third of Byzantine validators}\label{subsubsec:oneThirdByz} 

One may ask, why would Byzantine validators aim at going over the 1/3 threshold? Indeed, we just shown that Safety can be broken regardless of $\beta_0$; is it not the ultimate goal of Byzantine validators?
It is not obvious to determine what behaviour will harm the blockchain the most. We briefly discuss the impact Byzantine validators can have when they go over the 1/3 threshold in \autoref{subsec:revisitPBA}. We now examine what are the necessary conditions on $\beta_0$ and $p_0$ that permit Byzantine validator's stake to go over the one-third threshold.

The key ratio that translates into what we are looking for is the proportion of Byzantine validator's stake $\beta(t,p_0,\beta_0)$ over time: 
\begin{equation}
    \frac{\beta_0 e^{-3t^2/2^{28}} }{p_0(1-\beta_0) +(1-p)(1-\beta_0)e^{-t^2/2^{25}} + \beta_0 e^{-3t^2/2^{28}}}
\end{equation}

As expected at time $t=0$, $\beta(0,p_0,\beta_0)=\beta_0$. Now, let us investigate when this ratio is above the threshold of 1/3, i.e.:
\begin{equation}\label{eq:ByzOverOneThird}
\beta(t,p_0,\beta_0) \geq 1/3  
\end{equation}


The main difference with the previous scenario is that Byzantine validators seek to go over the 1/3 threshold, not to finalize quickly. This means that even after the proportion of honest active validators' stake and semi-active Byzantine validators' stake represent more than two-third of the stake on the branch, they do not finalize. Byzantine validators could finalize by staying active two epochs in a row, yet they do not do so in order to reach a higher stake proportion.

\

We construct a set containing the couples $(p_0,\beta_0)$ that can lead to $\beta$ to go over 1/3 (\autoref{eq:ByzOverOneThird}). To do so, we take the point reached by the ratio when the validators deemed inactive are ejected. This point gives the highest value reachable\footnote{There exist more values that can lead to go over one-third when considering a special corner case. If the Byzantine validators strategically finalize just before the expulsion of honest inactive validators, the decrease in inactivity penalties might not occur quickly enough to prevent the ejection of honest inactive validators.
In this particular scenario, Byzantine validators could potentially eject honest inactive participants while incurring fewer penalties themselves. This subtlety underscores the intricate dynamics at play during the inactivity leak.} for a particular $(p_0,\beta_0)$. For an intuition as to why this is the case, \autoref{fig:validatorsStake} let us visualize that the biggest gap between semi-active Byzantine stake and honest inactive stake is at the moment of expulsion of the honest inactive validators.
We have seen that inactive validators are ejected from the chain after 4685 epochs. We can thus evaluate the maximum ratio reachable $\beta_{\max}$ at time $t=4685$ when the inactive validators gets ejected: 
\begin{equation}
    \beta_{\max}(p_0,\beta_0) = \frac{\beta_0e^{-3\times (4685)^2/2^{28}}}{p_0(1-\beta_0) + \beta_0e^{-3\times (4685)^2/2^{28}}}.
\end{equation}

\begin{figure}[htbp]
    \vspace{-0.5cm}   \centerline{\includegraphics[scale=0.55]{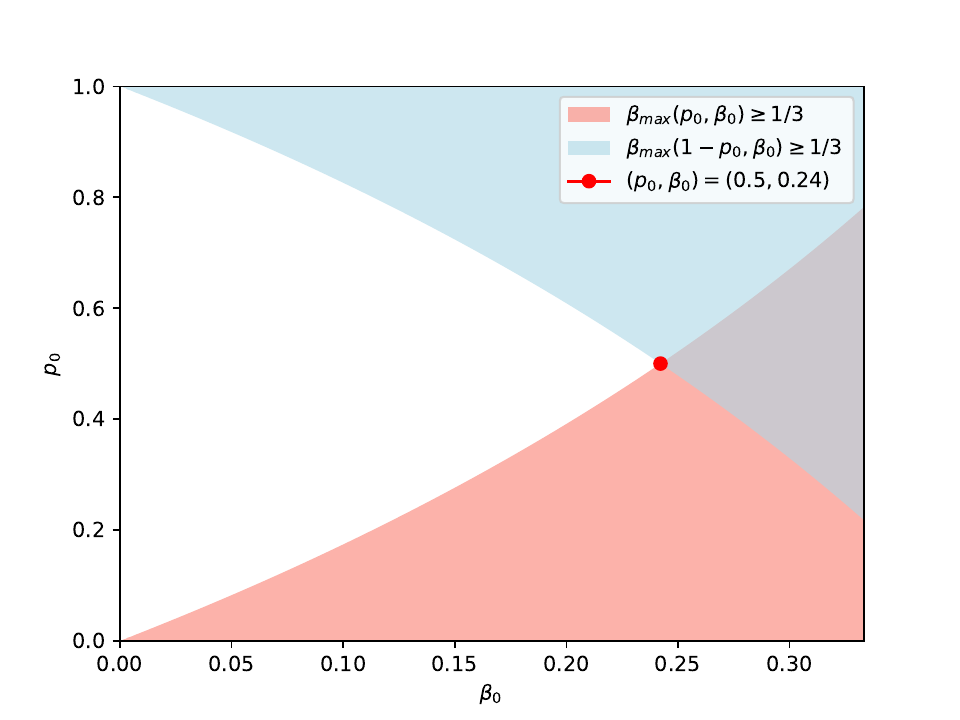}}
    \caption{Pairs $(p_0,\beta_0)$ such that $\beta_{\max}(p_0,\beta_0)\geq 1/3$. This figure gives a lower bound for which $(p_0,\beta_0)$ can result in the proportion of Byzantine validators exceeding 1/3 on both branches. } 
    \label{fig:pBetaShades}
\end{figure}

When this ratio is greater than 1/3, Byzantine validators have reached their goal. We show with \autoref{fig:pBetaShades} that Byzantine validators can actually go beyond the threshold of 1/3 on both branches simultaneously. The lower bound $\beta_0$ before this becomes possible is for $p_0=0.5$ when $\beta_0=1/(1+4e^{-3\times (4685)^2/2^{28}})=0.2421$. 

\textit{When the initial proportion of Byzantine validators is at least 0.2421, their proportion can eventually increase up to more than 1/3 of validators on both branches, exceeding the critical Safety threshold of voting power in each branch.}

Having explored scenarios in which protocol vulnerabilities manifest exclusively before \texttt{GST}, we now focus on potential threats posed by Byzantine validators after \texttt{GST}. Given the acknowledged impact of the \emph{Probabilistic Bouncing Attack} on Liveness \cite{pavloff_ethereum_2023}, our study extends to take the inactivity leak into account. 

\subsection{Revisiting the Probabilistic Bouncing attack}\label{subsec:revisitPBA}

This subsection revisits the Probabilistic Bouncing attack \cite{pavloff_ethereum_2023} showing that Byzantine validators could exceed the Safety threshold even during the synchronous period. Contrary to the previous scenarios, this one starts in the asynchronous period but unveils in the synchronous period. This demonstrates that the inactivity leak poses significant challenges even within the synchronous period, revealing its broader implication for blockchain security.

While analyzing the probabilistic bouncing attack \cite{pavloff_ethereum_2023}, the authors did not take into account the penalties, here, we fill this gap.

Let us note, that there is no problem with conflicting finalization as the attack is progressing after \texttt{GST}, in the synchronous period. In synchronous time, there is not enough delay for honest validators to miss a finalization on another branch. There would need to be more than two-third of the active stakes owned by Byzantine validators to break Safety in the synchronous period.

We briefly discussed the differences in gravity between conflicting finalization and having more than 1/3 of the stake owned by Byzantine validators. We left the actual comparison and the in-depth analysis of the gravity of going beyond the infamous threshold as a future work. 

\

We primarily focus on identifying specific scenarios that would disrupt the network. 
Thus, we give a detailed explanation of a scenario that could lead to Byzantine validators breaking the 1/3 threshold even during synchronous period (after \texttt{GST}).



Let us remind how the attack takes place.

\paragraph{Probabilistic Bouncing Attack \cite{pavloff_ethereum_2023}}

The attack can be summarized as follows: (1) A favorable setup  that partitions honest validators into two different views of the blockchain occurs. (2) At each epoch, Byzantine validators withhold their messages from honest validators, releasing them at the opportune time  to make some honest validators change their view.
(3) This attack continues as long as at least one Byzantine validators is proposer in the $j^{th}$ first slots of the epoch, where $j$ is a parameter of the protocol. The probability of the attack to continue for $k$ epochs with a proportion of $(1-\beta_0)$ honest validators is $(1-(1-\beta_0)^j)^k$.

We start by analyzing the outcome of a fork where a proportion $p_0$ of the \emph{honest} validators start on chain $A$ and $1-p_0$ of the honest validators start on chain $B$. 

We consider how would unfold a \textit{Probabilistic Bouncing Attack} taking the inactivity leak into account. A probabilistic bouncing attack lasting more than 4 epochs will necessarily cause an inactivity leak. Knowing this, we analyze the stakes of honest and Byzantine validators in this setting.

For this attack to continue, at each epoch, Byzantine validators cast their vote with a different chain as their candidate chain. They are active on each chain alternatively. Due to their inactivity every 2 epochs, they will get ejected of the chain after a total of 7653 epochs (4 weeks and 6 days).
Byzantine validators are active on each chain to ensure that justification only happens every two epoch, preventing finalization to occur. 

For this attack to continue indefinitely, Byzantine validators must ensure honest validators are split on the two branches according to two conditions.
A condition (a) that ensures that the honest validators are not enough to justify a chain on their own, and (b) that Byzantine validators can justify it afterwards with their withheld votes.
It means that (a) $p_0$ must not represent more than 2/3 of the stake, and (b) the proportions $p_0$ of honest validators and $\beta_0$ of Byzantine validators must represent more than two-thirds of the total stake.
The two necessary conditions are that (a) $p_0(1-\beta_0)<2/3$ and (b) $p_0(1-\beta_0)+\beta_0 > 2/3$. 
For the attack to function, we get that:
\begin{equation}\label{eq:p0conditions}
    \frac{2-3\beta_0}{3(1-\beta_0)} < p_0 < \frac{2}{3(1-\beta_0)} .
\end{equation}
We can see that the closer $\beta_0$ is to 0, the closer $p_0$ has to be from 2/3. This is to be expected as otherwise the Byzantine validators would be unable to justify the checkpoint with withheld votes. 

\begin{figure}[htbp]
    \centering
    \resizebox{0.5\columnwidth}{!}{
\begin{tikzpicture}[->,>=stealth,shorten >=1pt,node distance=2cm and 2.5cm]
  \tikzstyle{block} = [rectangle, draw, minimum width=1cm, minimum height=1cm]

  \node[block] (A) {A};
  \node[block, above right of=A] (B) {B};
  \node[block, below right of=A] (B') {B'};
  \node[block, right of= B] (C) {C};
  \node[block, right of= B'] (C') {C'};
  \node[block, right of= C] (D) {D};
  \node[block, right of= C'] (D') {D'};
  
  \draw (A) -- (B) node[midway, above, sloped]{$p_0$};
  \draw (A) -- (B') node[midway, below, sloped]{$1-p_0$}; 
  \draw (B) -- (C)  node[midway, above]{$1-p_0$} ;
  \draw (B') -- (C')  node[midway, below]{$p_0$} ;
  \draw (B) -- (C')  node[midway, below left, sloped]{$p_0$} ;
  \draw (B') -- (C)  node[midway, below right, sloped]{$1-p_0$}  ;
  
  \draw (C) -- (D)  node[midway, above]{$p_0$} ;
  \draw (C') -- (D')  node[midway, below]{$1-p_0$} ;
  \draw (C) -- (D')  node[midway, below left, sloped]{$1-p_0$} ;
  \draw (C') -- (D)  node[midway, below right, sloped]{$p_0$}  ;
  
\end{tikzpicture}
}
    \caption{This figure represents with a Markov chain the probability of an honest validator to change branch or not every epoch.
    During the attack, the Byzantine validators make sure that a proportion $p_0$ of honest validators are on one branch such that they can justify this branch later with their withheld votes (\autoref{eq:p0conditions}).}
    \label{fig:schemaAttack}
\end{figure}
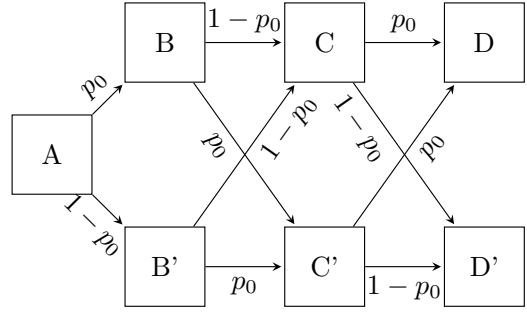

An illustration of an ongoing attack with the probability for honest validators to be on one chain or the other is depicted in \autoref{fig:schemaAttack}. We can see that at each epoch, a proportion $p_0$ is on one branch, whereas a proportion $1-p_0$ is on the other.

\paragraph*{Calculus} 

We are interested in the evolution of the proportion of Byzantine validators' stake $\beta$ during the attack. To look at the proportion, we analyze the evolution of the inactivity score over time for an honest validator randomly placed at each epoch. Looking at \autoref{fig:schemaAttack}, we see that after two epochs, there is a probability $p_0(1-p_0)$ to have been on branch $B$ for the two epochs, or on branch $A$ for the two epochs. 
The probability to have been on both branches  regardless of the order is $p_0^2+(1-p_0)^2$. 
From the point of view of a chain, the validators will be deemed inactive if they are active on the other chain. The probability of the inactivity score evolution after two epochs is the following:
\begin{equation}
    \left\{
    \begin{array}{ll}
        +8 :& p_0(1-p_0) \\
        +3 :& p_0^2+(1-p_0)^2 \\
        -2 :& p_0(1-p_0) \\
    \end{array}
\right.
\end{equation}

We can notice that the time-dependent probability of the inactivity score is the convolution of two random walks.  
The first random walk moves +4 with probability $p_0$ and -1 with probability $(1-p_0)$. The second is the opposite, it moves +4 with probability $(1-p_0)$ and -1 with probability $p_0$.

We place ourselves in the continuous case to be able to continue our analysis and find the stake of validators with the inactivity score distribution over time. To do so we use the fact that a random walk follows a Gaussian distribution when time is big using the central limit theorem. 
The expectation of the two random walks are $(5p_0-4t)$ and $(1-5p_0)t$, respectively, with both having a standard deviation of $25p_0(1-p_0)$. We disregard here the fact that the actual inactivity score is bounded by zero for analytical tractability. Allowing for negative values in the inactivity score can result in a reward instead of a penalty, which leads to a scenario conservatively estimating the loss of stake.
The convolution of these two random walks is the probability of the inactivity score $I$:
\begin{equation}
\phi(I,t)=\frac{1}{\sqrt{4\pi Dt}}\exp \left(-\frac{(I-Vt)^2}{4Dt}\right).
\end{equation}
With $D=25p_0(1-p_0)$ and $V=3/2$.
It now remains to find the distribution function of the stake $s$.  
We rewrite here the differential equation of the stake depending on $I$ previously described in \autoref{eq:stakeDerivative}:
\begin{equation}\label{eq:stakeDerivative}
    \frac{d s}{d t} = -\frac{I(t) s}{2^{26}}.
\end{equation}
Due to space constraints, we omit the detailed process for finding the distribution of $s$. The underlying intuition is that we can return to a known problem with a derivative being equal to a Brownian motion. By integrating the Brownian motion, we find our solution. 
This lead to the distribution of $s$ for any given time: 
\begin{equation}
    P(s,t)=\frac{2^{26}}{s\sqrt{\frac{4}{3}\pi Dt^3}}\exp\left(-\frac{(2^{26}\ln(s/32)+Vt^2/2)^2}{\frac{4}{3}Dt^3}\right).
\end{equation}



The stake follows a log normal distribution for which the cumulative function is:
\begin{equation}
    F(s,t)=\frac{1}{2}+\frac{1}{2} \operatorname{erf}\left[\frac{2^{26}\ln (s/32)+Vt^2/2}{\sqrt{\frac{4}{3}Dt^3}}\right]
\end{equation}

\begin{figure}[htbp]
    \centerline{\includegraphics[scale=0.55]{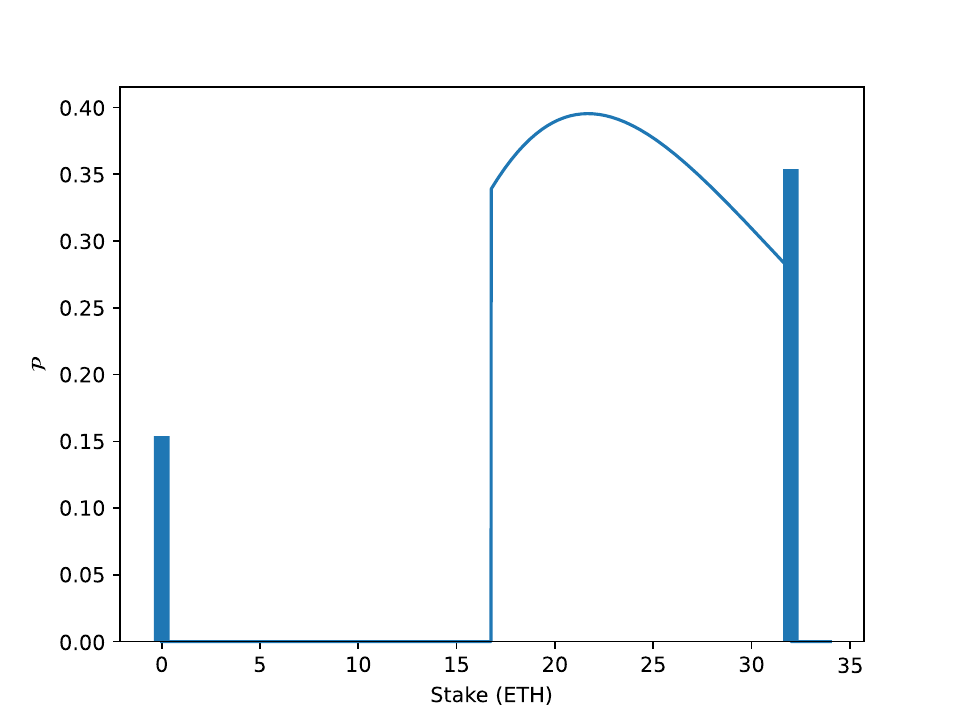}}
    \caption{This is a representation distribution of $\mathcal{P}$ at $t=4024$ with exaggerated standard deviation to give a better intuition of the distribution behavior.}
    \label{fig:mathcalP}
\end{figure}

Currently, the probability $P$ does not reflect the actual stake according to time since validators get ejected at 16.75 ETH and are stuck at 32 ETH. 
To emulate this mechanism, since we know the cumulative distribution function we can compute the new probability law $\mathcal{P}$:
\begin{equation}    
\mathcal{P}(x,t) = \begin{cases}
F(a,t) & \text{if } x = 0 \\ 
P(x,t) & \text{if } a < x < b \\
1-F(b,t) & \text{if }  x = b  
\end{cases}
\end{equation}
With $a=16.75$ and $b=32$. This new probability law takes into account the fact that if the stake is lower than 16.75 ETH it becomes 0,  and it is capped at 32 ETH.
The explicit expression of $\mathcal{P}$ reads:
\begin{equation}
\begin{split}
\mathcal{P}(x,t) = &\delta(x)\cdot F(a,t) + \delta(x-b)\cdot (1-F(b,t)) \\ 
&+ [H(x-a)\times H(b-x)] \cdot P(x,t),
\end{split}
\end{equation}
where $\delta$ is the Dirac distribution, and $H$ the Heaviside function. 
\autoref{fig:mathcalP} shows a visual representation of function $\mathcal{P}$.

The associated cumulative distribution function $\mathcal{F}$ of $\mathcal{P}$ is : 
\begin{equation}
\begin{split}
    \mathcal{F}(x,t) = \; &\int_{0}^{x} \mathcal{P}(s,t)ds \\
    = \; &F(a,t) + H(x-a)[F(x,t)-F(a,t)] \\
    &+ H(x-b)[1-F(x,t)] .
\end{split}
\end{equation}

With this we can evaluate the ratio of Byzantine validators and determine with what probability it will go beyond 1/3. 
We denote by $s_{\rm B}(t)$ the stake of a Byzantine validators and $s_{\rm H}(t)$ the stake of an honest validator.  We are looking for the probability such that
\begin{equation}
    \beta(t)=\frac{\beta_0s_{\rm B}(t)}{\beta_0s_{\rm B}(t)+(1-\beta_0)s_{\rm H}(t)}>\frac{1}{3},
\end{equation}
depending on the probability of $s_{\rm H}$ that we now know. This translates in:
\begin{equation}\label{eq:CDFratioByzOverOneThird}
    \mathcal{F}\left(\frac{2\beta_0}{1-\beta_0}s_{\rm B}(t),t\right),
\end{equation}
where $s_{\rm B}(t)$ the stake of a Byzantine validator follows the stake of a semi-active validator.

\begin{figure}[htbp]
\vspace{-0.5cm}
\centerline{\includegraphics[scale=0.55]{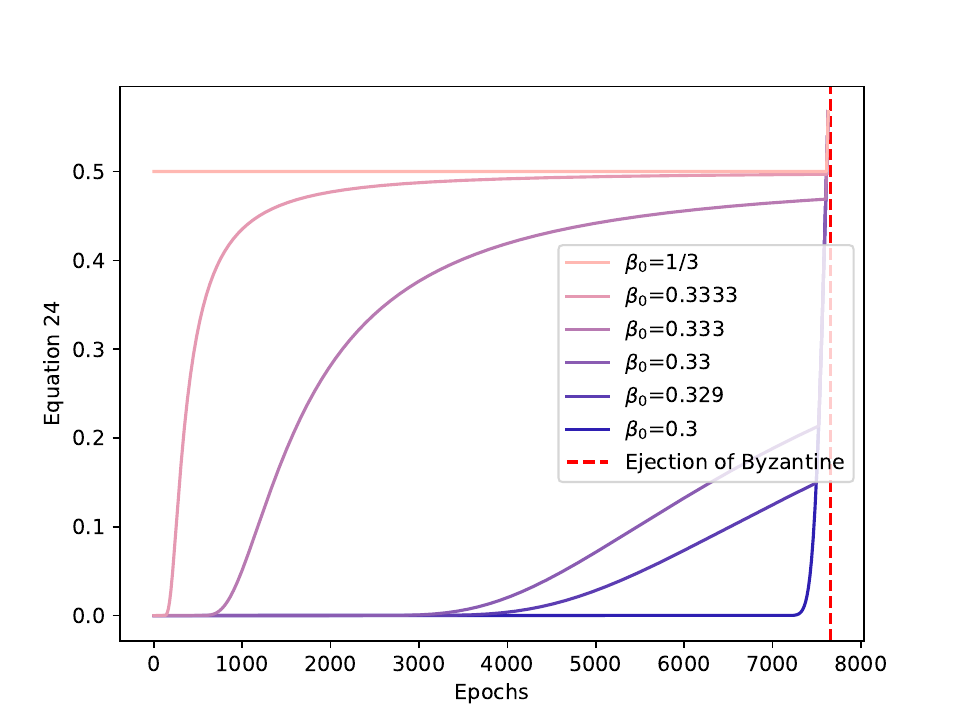}}
    \caption{We represent here \autoref{eq:CDFratioByzOverOneThird} according to time with various $\beta_0$.}
    \label{fig:CDFratioByzOverOneThird}
\end{figure}

We give a representation of \autoref{eq:CDFratioByzOverOneThird} for several values of $\beta_0$ with $p=0.5$ ($p_0$ does not have much impact on the curve as it just changes the variance slightly) in \autoref{fig:CDFratioByzOverOneThird}.

The figure gives insight as to how the proximity of $\beta_0$ to 1/3 can be harmful.
The explication of this phenomenon is due to the mean of the log-normal distribution being equivalent to $s_{\rm B}$ when $t$ is not too big. Looking at \autoref{eq:CDFratioByzOverOneThird} we can see that if $\beta_0=1/3$ then we are looking at  $\mathcal{F}(s_{\rm B}(t),t)$ which explains why its probability is 0.5.

The probability rises abruptly right before the expulsion of Byzantine validators, however, it is unlikely the probabilistic bouncing attack would last that long.
As an estimation, we can use the probability mentioned in \cite{pavloff_ethereum_2023} for an upper bound on the probability and to reach epoch 7000: $(1-(1-\beta_0)^8)^{7000}$ is equal to $1.01\times 10^{-121}$ for $\beta_0=1/3$. This negates all strategies of Byzantine validators that would need the probabilistic bouncing attack to last that long.

However, as \autoref{fig:CDFratioByzOverOneThird} shows, with $\beta_0$ nearing 1/3, Byzantine validators realistically have a high probability of quickly exceeding 1/3 of the stake, especially when considering the significant factor of the attack occurring on two branches.
Meaning that if a validator is active during an epoch on one branch, it is inactive on the other. Hence, the probability can be doubled for each curve. We can comprehend this by considering the case of $\beta_0=1/3$: after two epochs, the Byzantine validators have been active on each branch once.
If one branch has more validators that have been active on it for two epochs, the other branch will have honest validators incurring, on average, more penalties than the Byzantine validators. On this latter branch, the Byzantine stake will represent more than one-third of the total stake.

\textit{These results imply that, theoretically, within the synchronous period and with a proportion of Byzantine stake sufficiently close to 1/3 as well as a favorable initial setup, the probabilistic bouncing attack can pose a threat to the blockchain by allowing Byzantine validators to exceed the safety threshold of 1/3.} 

\section{Related Work}\label{sec:relatedWork}
  
While mechanisms similar to Ethereum's inactivity leak -- punishing the lack of validator activity -- exist elsewhere (e.g., \cite{wood_polkadot_2016, goodman_tezos_2014}), to the best of our knowledge, there has not been an analysis of the risk associated with potentially draining honest stake in a Byzantine-prone environment.

Initial efforts were made to intertwine the study of incentives with considerations of Liveness and Safety properties of the Ethereum protocol \cite{buterin_incentives_2020}. However, this early exploration discussed a preliminary version of the protocol \cite{buterin_casper_2017} and did not include an analysis of the inactivity leak. The most recent version of the protocol by its founder \cite{buterin_combining_2020} does not mention this mechanism. The inactivity leak still lacks a detailed examination, our work aims at filling this gap.

An investigation linking attestation penalties with the actions of Byzantine validators is presented in \cite{zhang_attestation_2023}. This work demonstrates how Byzantine validators can maliciously cause penalties for honest validators. Although similar in the enterprise, we differ since we focus on penalties predominant during the inactivity leak, i.e., the inactivity penalties and slashing, since during this period attestation penalties tend to be less significant.

On another line of research, some works study consensus protocols through the lens of game-theory (e.g., \cite{amoussou_guenou_et_al_2020, ABPT23, bias_et_al_2018, eyal_sirer_2018, fooladgar_incentive_2020, halpern_vilaca_2016, manshaei_et_al_2018, roughgarden_2020}), where all agents are rational, i.e, they behave strategically and have a maximization objective. This work differs from them on many aspects, first, we study the Ethereum PoS protocol, and second, instead of considering rational agents, we analyze the impact  classical agents considered in the distributed computing literature (correct and Byzantine) have on the Safety properties of the protocol under these incentive mechanisms.

\section{Discussion \& Conclusion}
This paper presents the first theoretical analysis of the inactivity leak, designed to restore finalization during catastrophic network failure. 
We point out situations where \textit{Byzantine actions expedite the loss of Safety, either with conflicting finalization or by increasing the Byzantine proportion over the one-third Safety threshold}. Interestingly, we 
showcase the possibility for Byzantine validators to exceed the one-third Safety threshold even during synchronous periods.



Our findings underscore the critical role of penalty mechanisms in BFT analysis. By shedding light on potential issues in the protocol design, we offer insights for future improvement and give tools to investigate them.


\section*{Acknowledgment}

We thank the DSN2024 reviewers for their detailed feedback and valuable suggestions, which helped improve our paper.

\bibliography{main}

\end{document}